# Competing redox and material redistribution during electrode ageing

Xiao Zhao[1,2,8], Yuchen Cao[3,8], Evan Z Carlson[1,2,8], Angel Burgos[1], Daniel Jacobs[3], Hanfeng Zhong[3], Lily Taylor[4], Haozhi Sha[3], Feng-yang Chen[1], Yu Shan[7], Hendrik Ohldag[2], Alexander Ditter[2], José A. Rodriguez[4], David Shapiro[2*], William Chueh[1,5,6*], and Jianwei Miao[3*]

*[1]Department of Materials Science and Engineering, Stanford University, Stanford, CA 94305, USA. [2]Advanced Light Source, Lawrence Berkeley National Laboratory, Berkeley, CA 94720, USA. [3]Department of Physics and Astronomy and California NanoSystems Institute, University of California, Los Angeles, CA 90095, USA. [4]Department of Chemistry and Biochemistry, University of California, Los Angeles, CA 90095, USA. [5]Applied Energy Division, SLAC National Accelerator Laboratory, Menlo Park, CA 94025, USA. [6]Department of Energy Science and Engineering, Stanford University, Stanford, CA 94305, USA. [7]Department of Materials Science and Engineering, University of California, Berkeley, CA 94720, USA.*

*[8]These authors contributed equally to this work.*

**Electrochemical processes can involve coupled redox reactions, active-material redistribution and morphological evolution at buried electrode–electrolyte interfaces[1–3]. Operando imaging has revealed nanoscale heterogeneity in these processes[3-9], but how reversible redox and cumulative material redistribution interact during cycles to govern electrode ageing remains unclear. Here we develop fast operando soft X-ray spectro-ptychography to track oxidation state, morphology and iron redistribution in the same nanoscale region of an alkaline iron anode over**

**20 h of cycling. We distinguish two coexisting processes operating on different timescales. Within each cycle, iron undergoes rapid, reversible conversion between metallic iron and iron oxyhydroxide (FeOOH) with little net iron redistribution. Across cycles, slower iron redistribution, consistent with dissolution and redeposition, promotes FeOOH-rich particle growth. Particle-resolved trajectories reveal that reversible conversion and iron accumulation coexist within individual particles, while particles with similar apparent growth exhibit markedly different changes in iron content. Local capacity initially increases as particles grow, then declines as more iron remains oxidized after charging and the reversible redox swing narrows. These findings suggest that cumulative Fe redistribution competes with reversible redox by progressively limiting the Fe available for charge storage. More generally, electrode ageing can begin during nearly reversible cycling, before capacity loss becomes apparent.**

Transition-metal electrodes store energy through redox reactions that can involve ion insertion, phase transformations and morphological changes[1,10-17]. Metal-containing species can also dissolve into the electrolyte and redeposit elsewhere, redistributing active material and reorganizing reaction sites[3,18]. These processes coexist, with local redox sustaining reversible charge storage within each cycle while cumulative metal redistribution can reduce electrode capacity over successive cycles[3,17,18]. Operando studies have revealed heterogeneous redox and phase transformations[4,5,7,9,16,19], particle- and subparticle-scale reaction dynamics[6,8,20-22], lattice strain and electrode degradation[9,23-25], and active-material migration[3,26], often focusing on selected electrochemical states or a limited number of cycles. However, how reversible local redox and cumulative metal redistribution evolve within the same nanoscale region and contribute to electrode ageing remains unresolved.

Addressing this question requires operando measurements that combine chemical-state sensitivity, nanoscale resolution and rapid acquisition with long-term stability in a liquid electrochemical cell. This combination is particularly challenging in aqueous systems, where electrode reactions are strongly coupled to the electrolyte chemistry[21,27–31]. Soft X-ray spectro-ptychography combines X-ray absorption spectroscopy with coherent diffractive imaging to map chemical states and morphology at nanoscale resolution[32-36]. Previous studies have revealed chemical heterogeneity in battery materials ex situ[37,38] and tracked chemical and morphological changes during electrocatalysis and electrodeposition[39,40]. Extending these capabilities to electrode ageing requires repeated chemical imaging of the same region, fast enough to capture reactions within individual cycles and sustained long enough to follow cumulative changes across many cycles. Here we develop fast operando soft X-ray spectro-ptychography to investigate redox and cumulative material redistribution during electrode ageing. Using an alkaline iron anode as a model system, we distinguish reversible chemical changes from the gradual redistribution of iron and reveal how continued particle growth is linked to incomplete reduction and declining charge storage.

## Fast operando soft X-ray spectro-ptychography

To track nanoscale changes in electrode morphology and chemical state during cycling, we developed fast operando soft X-ray spectro-ptychography incorporating a liquid electrochemical flow cell (Fig. 1a,b and Methods). The cell confined a 3 μm-thick layer of argon-saturated 0.1 M potassium hydroxide (Ar-saturated KOH) between two X-ray-transparent silicon nitride windows. One window carried a tin-doped indium oxide current collector supporting 25-nm-thick Fe microelectrodes with lateral dimensions of 10 μm × 35 μm. Ptychographic images were reconstructed from coherent X-ray diffraction patterns recorded at overlapping probe positions[33,34] (Extended Data Fig. 1). To obtain chemical contrast, we selected three photon energies near the Fe $L_3$ absorption edge (701.0, 707.6 and 709.0 eV), guided by operando X-ray absorption spectra of the electrode in its charged and discharged states (Fig. 1c). These energies sample the pre-edge background, the metallic-Fe resonance and the $Fe^{3+}$ resonance respectively[41,42]. Sequential scans of the same region at these energies were reconstructed into three-energy image stacks, from which we derived maps of Fe oxidation state, morphology and Fe content per unit area during charging and discharging (Methods).

To acquire these stacks rapidly enough to resolve within-cycle changes, we used a highly defocused probe that enabled a sparser scan grid while maintaining sufficient overlap for reliable reconstruction (Extended Data Fig. 2 and Methods). The reduced number of exposures increased imaging speed by a factor of 6.25 and reduced the estimated radiation dose per scan by the same factor under matched conditions. This strategy achieved a single-energy imaging rate of up to 14.6 $\mu m^2\ min^{-1}$ with a full-period spatial resolution of 34 nm (Extended Data Figs. 2b and 3). Each three-energy stack required 3.3–3.6 min, providing about 10 chemically resolved time points per typical 35-min electrochemical cycle. Repeated imaging yielded 311 three-energy stacks from the same 4 μm × 4 μm region over 20 h, capturing both within-cycle responses and cumulative particle-level changes as the electrode evolved from reversible cycling to degradation (Extended Data Fig. 4 and Supplementary Videos 1 and 2).

We applied this platform to a thin-film iron anode cycled in Ar-saturated 0.1 M KOH, which undergoes initially reversible Fe–FeOOH conversion alongside progressive morphological restructuring and eventual capacity fade[43–47]. Cyclic voltammetry showed two redox peak pairs near −0.25/−0.10 V and −0.15/0.33 V versus the reversible hydrogen electrode (Fig. 1d). Separate electrochemical measurements on similarly prepared Fe films revealed three stages: initial formation during cycles 1–5, relatively stable capacity through cycle 35, and rapid capacity fade thereafter (Extended Data Fig. 5b). Ex situ Raman spectroscopy, electron microscopy and atomic force microscopy revealed progressive roughening of the Fe film and growth of vertically oriented FeOOH nanoplates (Fig. 1e and Extended Data Fig. 5c–e). Post-mortem electron microscopy of the operando-imaged region showed a similar nanoplate morphology (Extended Data Fig. 6b,c).

**Rapid local conversion sustains reversible charge storage**

We first examine cycles 8–14, during which the Fe anode exhibited nearly reversible charge–discharge behaviour (Fig. 2a,b and Supplementary Video 1). From each three-energy stack, we derive pixel-resolved maps of Fe oxidation state, absorption-derived equivalent thickness (i.e. fitted topography) and equivalent Fe areal mass density, which provides an estimate of the local Fe content per unit area (Methods). The spatially averaged oxidation state closely follows the applied potential and current, while the fitted topography varies reversibly within each cycle. These changes reflect reversible conversion between states dominated by metallic Fe and FeOOH, accompanied by local expansion and contraction.

The height-rendered maps in Fig. 2c show that this response is strongly heterogeneous across the Fe film. During discharge, FeOOH-rich regions become more oxidized and increase in fitted thickness, with oxidation states approaching +3. During the subsequent charge, these features are reduced and shrink as the film returns towards the metallic state. Larger features remain oxidized for longer, consistent with slower local reduction kinetics. By the end of charge, however, most of the imaged region returns close to the metallic state. The repeated growth and shrinkage of FeOOH-rich regions at the same locations therefore reflects a largely reversible local chemo-morphological response within each cycle, despite pronounced particle-to-particle differences. The equivalent Fe areal-mass maps provide the key distinction between conversion and dissolution-redeposition. Although oxidation state and fitted topography change within each cycle, the local Fe content remains nearly constant across the imaged region (Fig. 2c and Supplementary Video 1). This contrast indicates that the dominant intra-cycle response is local conversion coupled to uptake and release of electrolyte-derived ion species, with little net Fe redistribution. The electrode can therefore undergo substantial chemical and thickness changes while largely preserving its local Fe inventory.

Figure 2d compares the charged and discharged states during cycle 13. Particle-rich regions exhibit the largest changes in fitted topography and Fe oxidation state, together with the highest local discharge capacity, calculated from changes in effective FeOOH thickness (Methods). These largely reversible chemical and thickness changes, with little net change in local Fe content, support local conversion coupled to uptake and release of electrolyte species, likely $OH^-$(Fig. 2e). The three-energy measurements do not identify these species or resolve short-lived intermediates involved in alkaline Fe redox chemistry[43-47]. This reversible within-cycle response provides a baseline for identifying cumulative Fe redistribution over successive cycles.

**Progressive FeOOH particle growth drives capacity loss**

We next examine cycles 16–25, which span the transition from reversible cycling to capacity fade (Fig. 3a,b and Supplementary Video 2). Intra-cycle redox oscillations persist, but the spatially averaged Fe oxidation state progressively increases in both the charged and discharged states. Consequently, an increasing fraction of Fe remains oxidized at the end of charge rather than returning towards metallic Fe. This accumulating residual oxidation marks a growing irreversible component of the reaction and coincides with the onset of capacity loss. The electrode therefore degrades through increasingly incomplete reduction, even as intra-cycle conversion continues within each cycle.

The height-rendered maps in Fig. 3c, Extended Data Fig. 7 and Supplementary Video 2 reveal how degradation evolves spatially over successive cycles. FeOOH-rich regions progressively increase in fitted thickness and become more heterogeneous, consistent with cumulative particle growth and morphological restructuring. Unlike during the earlier reversible regime, the corresponding charged states no longer return close to a uniform metallic state. Instead, persistent FeOOH-rich regions become increasingly difficult to reduce as cycling proceeds. Characterization of the imaged region after cycling shows a similarly roughened surface covered with dense FeOOH-rich features, consistent with the operando observations (Extended Data Fig. 6).

Figure 3c further shows how continued particle growth progressively limits reversible charge storage. From cycles 16 to 19, the intra-cycle swings in fitted topography and oxidation state increase slightly, and the local discharge capacity becomes concentrated in FeOOH-rich regions. With further cycling, the particles continue to grow (Extended Data Fig. 7 and Supplementary Video 2), but the thickness change, oxidation-state swing and local capacity decline after cycle 19. Particle growth therefore initially increases local charge storage, whereas continued growth leaves an increasing fraction of Fe oxidized after charge, limits complete reduction and narrows the reversible redox swing. These observations are consistent with increasing electron-transport limitations in growing FeOOH-rich particles, hindering their complete reduction. The concurrent decrease in hydrogen-evolution current further suggests a less metallic electrode surface (Fig. 3a). Together, these changes link persistent FeOOH-rich particle growth to declining electrochemical reversibility and progressive capacity loss.

**Particle-dependent growth pathways connect reversible redox to electrode ageing**

To examine how individual particles evolve over successive cycles, we tracked 79 particles in their discharged states during cycles 8–14 (Extended Data Fig. 8). Figure 4a shows 31 representative particle trajectories, with capacity plotted against cycle number and marker size and colour indicating particle volume and mean Fe oxidation state, respectively. Across all 79 particles, volume and capacity are strongly correlated, whereas mean Fe oxidation state correlates more weakly with either (Extended Data Fig. 9). However, similar increases in apparent volume and capacity cannot distinguish redox-related expansion from Fe accumulation.

Particles 1 and 2 illustrate this distinction (Fig. 4b,c,e). During cycle 13, particle 1 exhibits a reversible change in fitted thickness of approximately 50% with little net change in Fe content, consistent with local conversion. Across cycles 8–14, however, both its fitted thickness and Fe content progressively increase, indicating cumulative Fe accumulation. Particle 2 also thickens over successive cycles but gains little net Fe, despite pronounced changes in oxidation state. Its apparent growth is therefore more consistent with redox-related expansion than with Fe accumulation.

The particle trajectories reveal a continuum of growth pathways involving local conversion and Fe redistribution (Fig. 4d). Approximately 84% of the analysed growing particles gain both fitted thickness and Fe content, whereas the remaining 16% thicken with little net Fe accumulation (Fig. 4d and Extended Data Fig. 10). Importantly, cumulative Fe redistribution is already evident during nearly reversible cycling, before the onset of capacity fade. Such redistribution is consistent with dissolution of Fe-containing species into the alkaline electrolyte followed by redeposition onto FeOOH-rich regions[3,48].

These observations support the mechanism summarized in Fig. 4f. Local conversion sustains reversible charge storage within each cycle, while slower Fe redistribution contributes to particle growth across cycles. Particle growth initially accompanies increasing imaging-derived local capacity, but continued growth is associated with persistent FeOOH-rich regions, incomplete reduction and declining capacity. These findings suggest that cumulative Fe redistribution progressively concentrates Fe in regions that resist reduction, limiting its availability for reversible conversion and giving the two processes competing effects on long-term charge storage.

## Conclusion

Fast operando soft X-ray spectro-ptychography enables sustained nanoscale measurements of chemical state, morphology and active-material redistribution in the same electrode region during cycling. Our findings reveal that cumulative Fe redistribution begins during nearly reversible cycling, before the decline in imaging-derived local capacity. With continued cycling, FeOOH-rich regions grow, reduction becomes increasingly incomplete and reversible charge storage declines. This evolution supports a mechanism in which slow Fe redistribution progressively limits the active material available for rapid reversible conversion, providing a physical link between cumulative material evolution and electrode ageing.

The interplay between reversible redox and cumulative material redistribution has implications beyond alkaline iron electrodes. Across batteries, electrocatalysis, corrosion and electrodeposition, material transport and structural reorganization can alter reaction sites, local kinetics and the accessibility of active material[3,5,26,40,48-50]. Long-duration, same-region chemical imaging provides a means to determine how these cumulative changes influence electrochemical function and guide strategies to improve long-term stability.

**References**

1. Sood, A. et al. Electrochemical ion insertion from the atomic to the device scale. *Nat. Rev. Mater*. **6**, 847–867 (2021).
2. Ebner, M., Marone, F., Stampanoni, M. & Wood, V. Visualization and quantification of electrochemical and mechanical degradation in Li ion batteries. *Science* **342**, 716–720 (2013).
3. Zhang, Y. et al. Operando characterization and regulation of metal dissolution and redeposition dynamics near battery electrode surface. *Nat. Nanotechnol*. **18**, 790–797 (2023).
4. Lim, J. et al. Origin and hysteresis of lithium compositional spatiodynamics within battery primary particles. *Science* **353**, 566–571 (2016).
5. Mefford, J. T. et al. Correlative operando microscopy of oxygen evolution electrocatalysts. *Nature* **593**, 67–73 (2021).
6. Li, J. et al. Dynamics of particle network in composite battery cathodes. *Science* **376**, 517–521 (2022).
7. Dai, H. et al. Unraveling chemical origins of dendrite formation in zinc-ion batteries via in situ/operando X-ray spectroscopy and imaging. *Nat. Commun*. **15**, 8577 (2024).

8. Merryweather, A. J. et al. Operando monitoring of single-particle kinetic state-of-charge heterogeneities and cracking in high-rate Li-ion anodes. *Nat. Mater.* **21**, 1306–1313 (2022).
9. Lee, D. et al. Strain-associated nanoscale fluctuating lithium transport within single-crystalline LiNi1/3Mn1/3Co1/3O2 cathode particles. *Nat. Commun.* **16**, 9018 (2025).
10. Wang, H. et al. Recent advances in conversion-type electrode materials for post lithium-ion batteries. *ACS Mater. Lett.* **3**, 956–977 (2021).
11. Heo, J. et al. Amorphous iron fluorosulfate as a high-capacity cathode utilizing combined intercalation and conversion reactions with unexpectedly high reversibility. *Nat. Energy* **8**, 30–39 (2023).
12. Pan, H. et al. Reversible aqueous zinc/manganese oxide energy storage from conversion reactions. *Nat. Energy* **1**, 16039 (2016).
13. Sun, Y. et al. Solvent co-intercalation in layered cathode active materials for sodium-ion batteries. *Nat. Mater.* **24**, 1441–1449 (2025).
14. Guo, F. et al. Revitalizing iron redox by anion-insertion-assisted ferro- and ferri-hydroxides conversion at low alkalinity. *J. Am. Chem. Soc.* **144**, 11938–11942 (2022).
15. Jagadeesan, S. N. et al. Chloride insertion enhances the electrochemical oxidation of iron hydroxide double-layer hydroxide into oxyhydroxide in alkaline iron batteries. *Chem. Mater.* **35**, 6517–6526 (2023).
16. Li, L. et al. Visualization of electrochemically driven solid-state phase transformations using operando hard X-ray spectro-imaging. *Nat. Commun.* **6**, 6883 (2015).
17. Jin, H. et al. Switching from insertion to conversion for multielectron aqueous vanadium batteries. *Nat. Mater.* **25**, 1590–1598 (2026).
18. Promi, A. T. et al. Spatiotemporal and statistical mapping of transition metal equilibria in alkaline media. *Chem. Mater.* **38**, 1315–1324 (2026).

19. Sasaki, Y. et al. Development of Operando Hard X-ray Ptychography: Application to Thin-Film All-Solid-State Lithium-Ion Batteries. *J. Phys. Chem. C* **129**, 10624–10632 (2025)

20. Han, J. et al. Avalanche-like intercalation and intraparticle correlations in graphite. *Nature* **656**, 364–369 (2026).

21. Zhang, W. et al. Optical nanoscopy of spatiotemporal metal stripping cooperativity at single-ion and subparticle resolution. *Nat. Mater*. **25**, 782–790 (2026).

22. Peng, Y. et al. Operando X-ray imaging reveals size-dependent evolution of cobalt oxide thermochemical material during thermal redox cycles. *Nat. Commun*. **16**, 11278 (2025).

23. Sun, T. et al. Electrode strain dynamics in layered intercalation battery cathodes. *Science* **390**, 1272–1277 (2025).

24. Liu, T. et al. Origin of structural degradation in Li-rich layered oxide cathode. *Nature* **606**, 305–312 (2022).

25. Kimura, Y. et al. 5D Analysis of Capacity Degradation in Battery Electrodes Enabled by Operando CT-XANES. *Small Methods* **7**, 2300310 (2023).

26. Yang, Y. et al. Operando probing dynamic migration of copper carbonyl during electrocatalytic $CO_2$ reduction. *Nat. Catal*. **8**, 579–594 (2025).

27. Luo, H. et al. Aqueous iron-ions batteries: status, solutions, and prospects. *Adv. Mater*. **37**, 2507978 (2025).

28. Liang, Y. & Yao, Y. Designing modern aqueous batteries. *Nat. Rev. Mater*. **8**, 109–122 (2023).

29. Yuan, Y. et al. Understanding intercalation chemistry for sustainable aqueous zinc–manganese dioxide batteries. *Nat. Sustain*. **5**, 890–898 (2022).

30. Bi, S., Wang, S., Yue, F., Tie, Z. & Niu, Z. A rechargeable aqueous manganese-ion battery based on intercalation chemistry. *Nat. Commun*. **12**, 6991 (2021).

31. Kankanallu, Varun R., et al. Elucidating a dissolution–deposition reaction mechanism by multimodal synchrotron X-ray characterization in aqueous Zn/MnO2 batteries. *Energy Environ. Sci.* **16 (6)**, 2464-2482 (2023).
32. Miao, J. Computational microscopy with coherent diffractive imaging and ptychography. *Nature* **637**, 281–295 (2025).
33. Urquhart, S. G. X-ray spectroptychography. *ACS Omega* **7**, 11521–11529 (2022).
34. Shapiro, D. A. et al. Chemical composition mapping with nanometre resolution by soft X-ray microscopy. *Nat. Photon*. **8**, 765–769 (2014).
35. Hitchcock, A. P. et al. Comparison of soft X-ray spectro-ptychography and scanning transmission X-ray microscopy. *J. Electron Spectrosc. Relat. Phenom*. **276**, 147487 (2024).
36. Shapiro, D. A. et al. An ultrahigh-resolution soft X-ray microscope for quantitative analysis of chemically heterogeneous nanomaterials. *Sci. Adv*. **6**, eabc4904 (2020).
37. Yu, Y.-S. et al. Three-dimensional localization of nanoscale battery reactions using soft X-ray tomography. *Nat. Commun*. **9**, 921 (2018).
38. Sun, T. et al. Soft X-ray ptychography chemical imaging of degradation in a composite surface-reconstructed Li-rich cathode. *ACS Nano* **15**, 1475–1485 (2021).
39. Zhang, C. et al. Copper carbon dioxide reduction electrocatalysts studied by in situ soft X-ray spectro-ptychography. *Cell Rep. Phys. Sci*. **4**, 101665 (2023).
40. Bozzini, B. et al. Soft X-ray ptychography as a tool for in operando morphochemical studies of electrodeposition processes with nanometric lateral resolution. *J. Electron Spectrosc. Relat. Phenom*. **220**, 147–155 (2017).
41. Sassi, M., Pearce, C. I., Bagus, P. S., Arenholz, E. & Rosso, K. M. First-principles Fe $L_{2,3}$-edge and O K-edge XANES and XMCD spectra for iron oxides. *J. Phys. Chem. A* **121**, 7613–7618 (2017).

42. Miedema, P. S. & de Groot, F. M. F. The iron L edges: Fe 2p X-ray absorption and electron energy loss spectroscopy. *J. Electron Spectrosc. Relat. Phenom.* **187**, 32–48 (2013).

43. Weinrich, H. et al. Understanding the nanoscale redox-behavior of iron-anodes for rechargeable iron-air batteries. *Nano Energy* **41**, 706–716 (2017).

44. Lee, D.-C., Lei, D. & Yushin, G. Morphology and phase changes in iron anodes affecting their capacity and stability in rechargeable alkaline batteries. *ACS Energy Lett.* **3**, 794–801 (2018).

45. Tan, S. F. et al. Electrochemical reactivity and stability of the Fe electrode in alkaline electrolyte. *Adv. Funct. Mater.* **35**, 2407561 (2025).

46. He, Z. et al. Iron metal anode for aqueous rechargeable batteries. *Mater. Today Adv.* **11**, 100156 (2021).

47. Weninger, B. M. H., Thijs, M. A., Nijman, J. A. C., Van Eijck, L. & Mulder, F. M. Neutron diffraction study of a sintered iron electrode in operando. *J. Phys. Chem. C* **125**, 16391–16402 (2021).

48. Dražić, D. M. & Hao, C. S. The anodic dissolution process on active iron in alkaline solutions. *Electrochim. Acta* **27**, 1409–1415 (1982).

49. Chung, D. Y. et al. Dynamic stability of active sites in hydr(oxy)oxides for the oxygen evolution reaction. *Nat. Energy* **5**, 222–230 (2020).

50. Kuai, C. et al. Phase segregation reversibility in mixed-metal hydroxide water oxidation catalysts. *Nat. Catal.* **3**, 743–753 (2020).

## Figures and figure legends

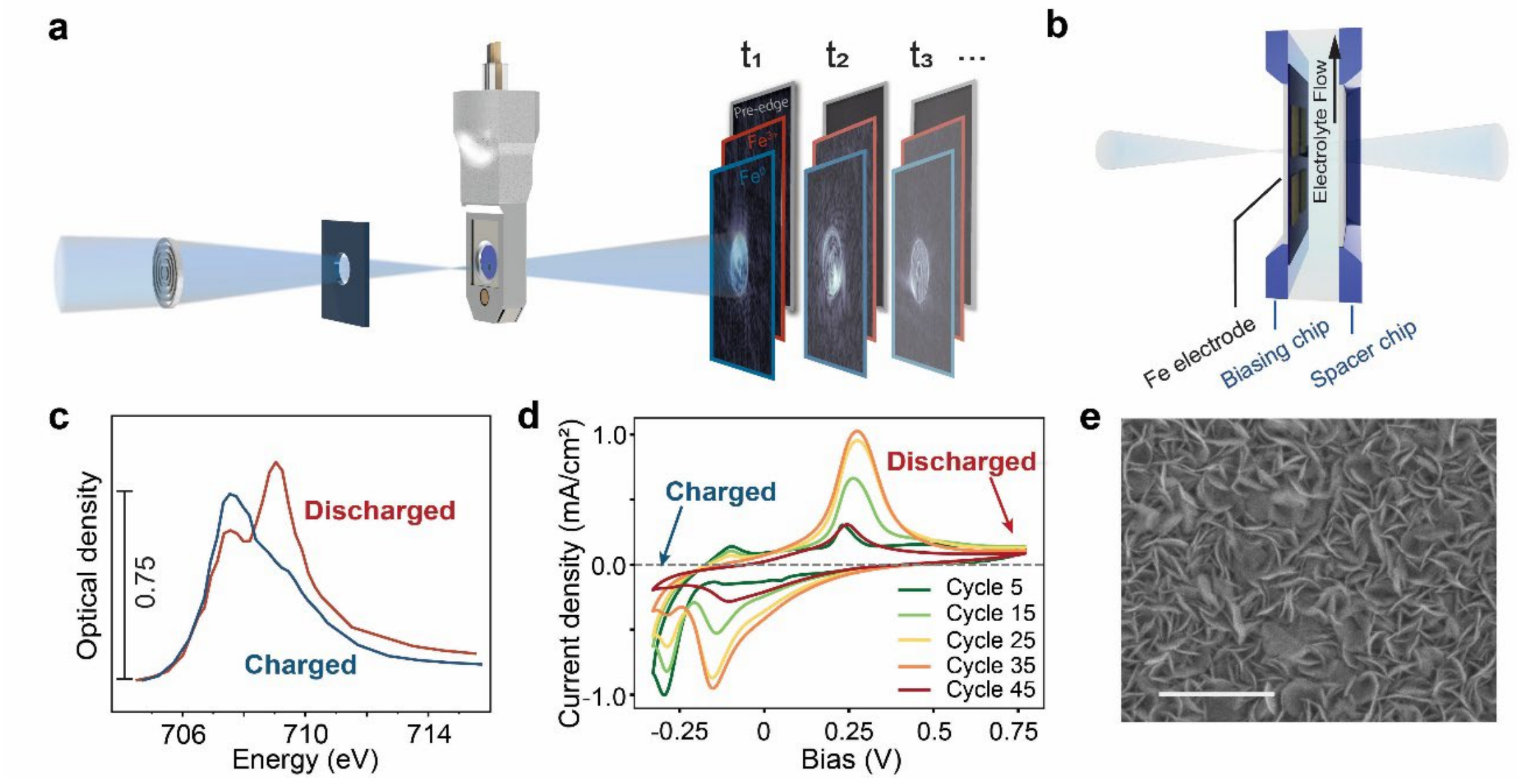

**Fig. 1 | Fast operando soft X-ray spectro-ptychography**. **a,** Experimental schematic. A highly defocused soft X-ray probe scans an Fe anode on a sparse grid with sufficient overlap for ptychographic reconstruction. Ptychographic datasets, each comprising diffraction patterns from multiple scan positions, are acquired sequentially at three photon energies and repeatedly during electrochemical cycling. **b,** Enlarged view of the liquid electrochemical flow cell. Two X-ray-transparent silicon nitride windows enclose a 3 µm-thick electrolyte channel. The electrode chip carries Fe thin-film microelectrodes on an indium tin oxide current collector and a Pt counter electrode. The external Ag/AgCl reference electrode is not shown. **c,** Operando Fe $L_3$-edge X-ray absorption spectra of the same electrode in the charged (blue) and discharged

(red) states, averaged over the imaged region. **d,** Cyclic voltammograms in argon-saturated 0.1 M KOH for cycles 5, 15, 25, 35 and 45. Potentials are referenced to the reversible hydrogen electrode. Cycling performance is shown in Extended Data Fig. 5a,b. **e,** Scanning electron microscopy image of the Fe anode after cycling in 0.1 M KOH, showing platelet-like FeOOH features. Scale bar, 1 µm.

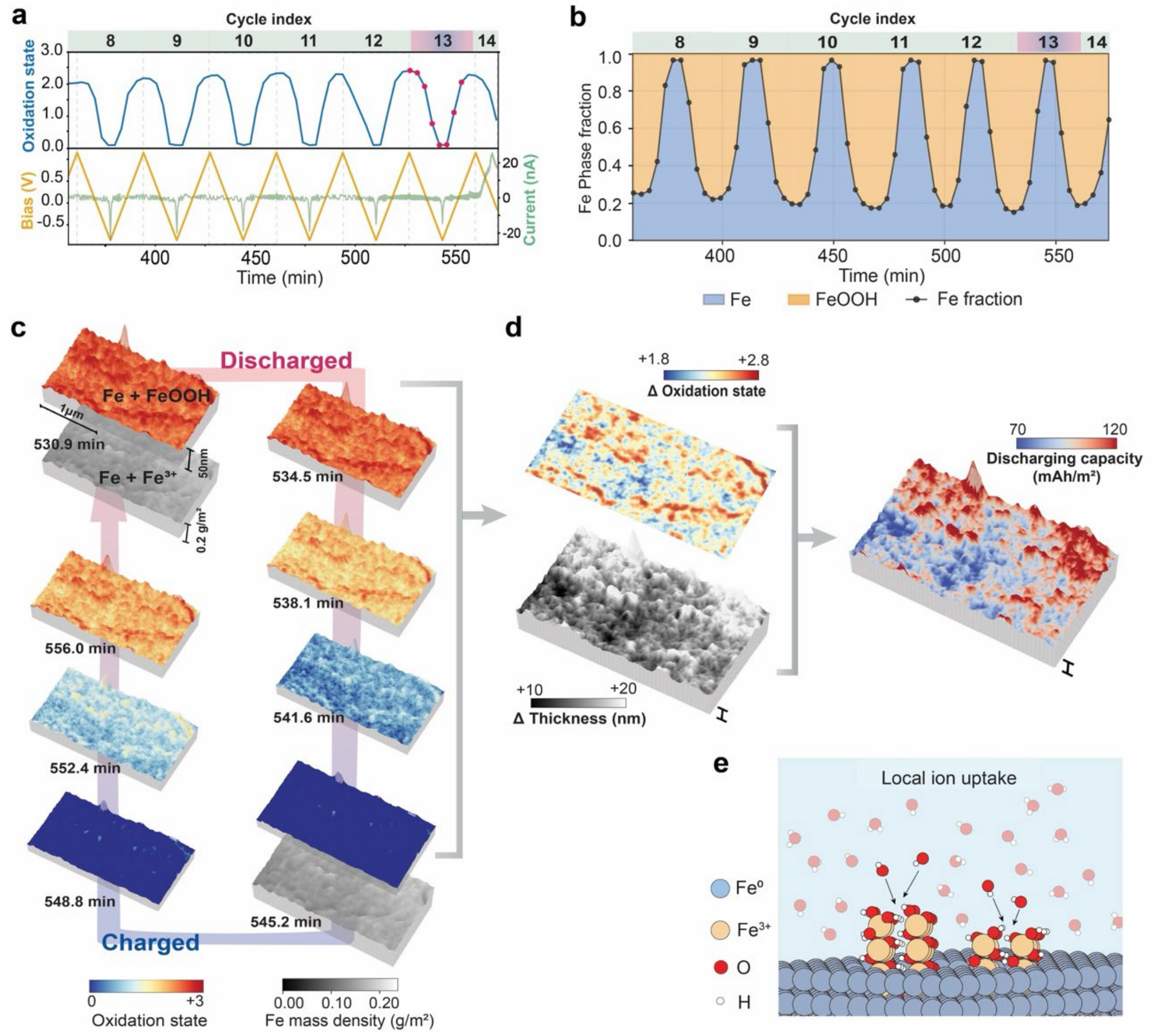

**Fig. 2 | Reversible local redox in an alkaline Fe anode. a**, Spatially averaged Fe oxidation state (blue, top) during cycles 8–14, with applied potential (yellow) and current (green) below. Magenta markers identify time points in **c**. **b,** Estimated metallic-Fe (blue) and FeOOH (orange) fractions from spectral fitting. Black points and line indicate the metallic-Fe fraction. **c,** Selected height-rendered maps from cycle 13. Height represents absorption-derived

equivalent thickness (fitted topography), and colour indicates Fe oxidation state. Grey surfaces show equivalent Fe areal mass density in the discharged and charged states. FeOOH-rich features thicken during discharge and shrink during charge, with little detectable change in local Fe content. **d,** Discharged-minus-charged changes in fitted topography (lower left) and Fe oxidation state (upper left) during cycle 13, with derived local discharge capacity (right), calculated from the effective FeOOH thickness change (Methods). In the topography-change and capacity maps, height represents fitted topography and colour indicates the change or capacity. **e,** Schematic of local ion uptake coupled to Fe redox, without assigning a specific electrolyte-derived species.

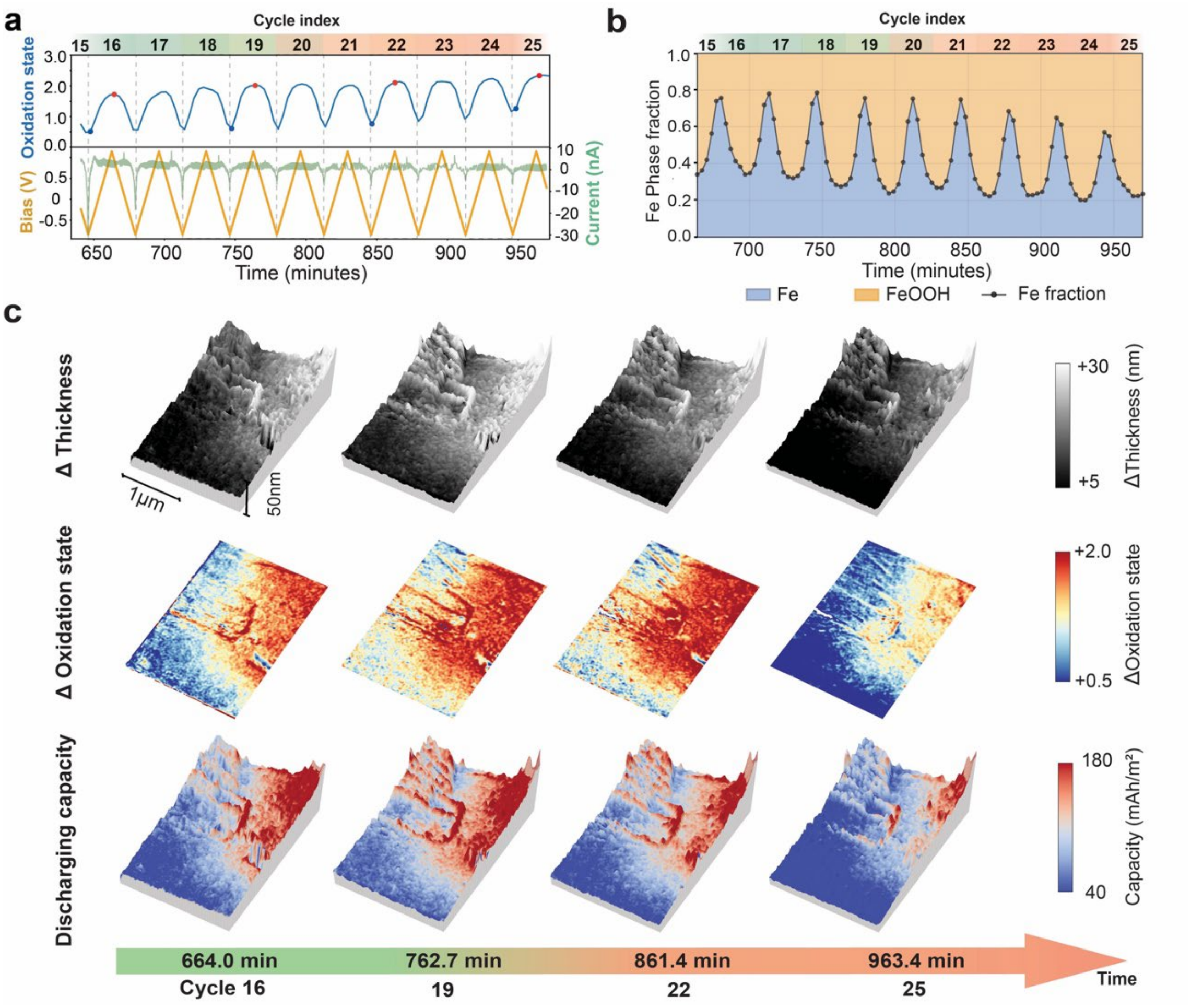

**Fig. 3 | Progressive FeOOH particle growth drives capacity loss. a**, Spatially averaged Fe oxidation state (blue, top) during cycles 16–25, with applied potential (orange) and current

(green) below. Blue and red markers identify the charged and discharged states used in **c**, respectively. **b,** Estimated metallic-Fe (blue) and FeOOH (orange) fractions from spectral fitting. Black points and line indicate the metallic-Fe fraction. Cyclic Fe–FeOOH conversion persists, but increasing residual FeOOH after charge indicates progressively incomplete reduction. **c,** Discharged-minus-charged changes in fitted topography (top) and Fe oxidation state (middle), with derived local discharge capacity (bottom), for cycles 16, 19, 22 and 25 from left to right. In the top and bottom rows, height represents absorption-derived equivalent thickness (fitted topography), and colour indicates thickness change or capacity. Capacity is calculated from the effective FeOOH thickness change (Methods). The three measures increase from cycle 16 to 19, then decline by cycle 25 despite continued particle growth.

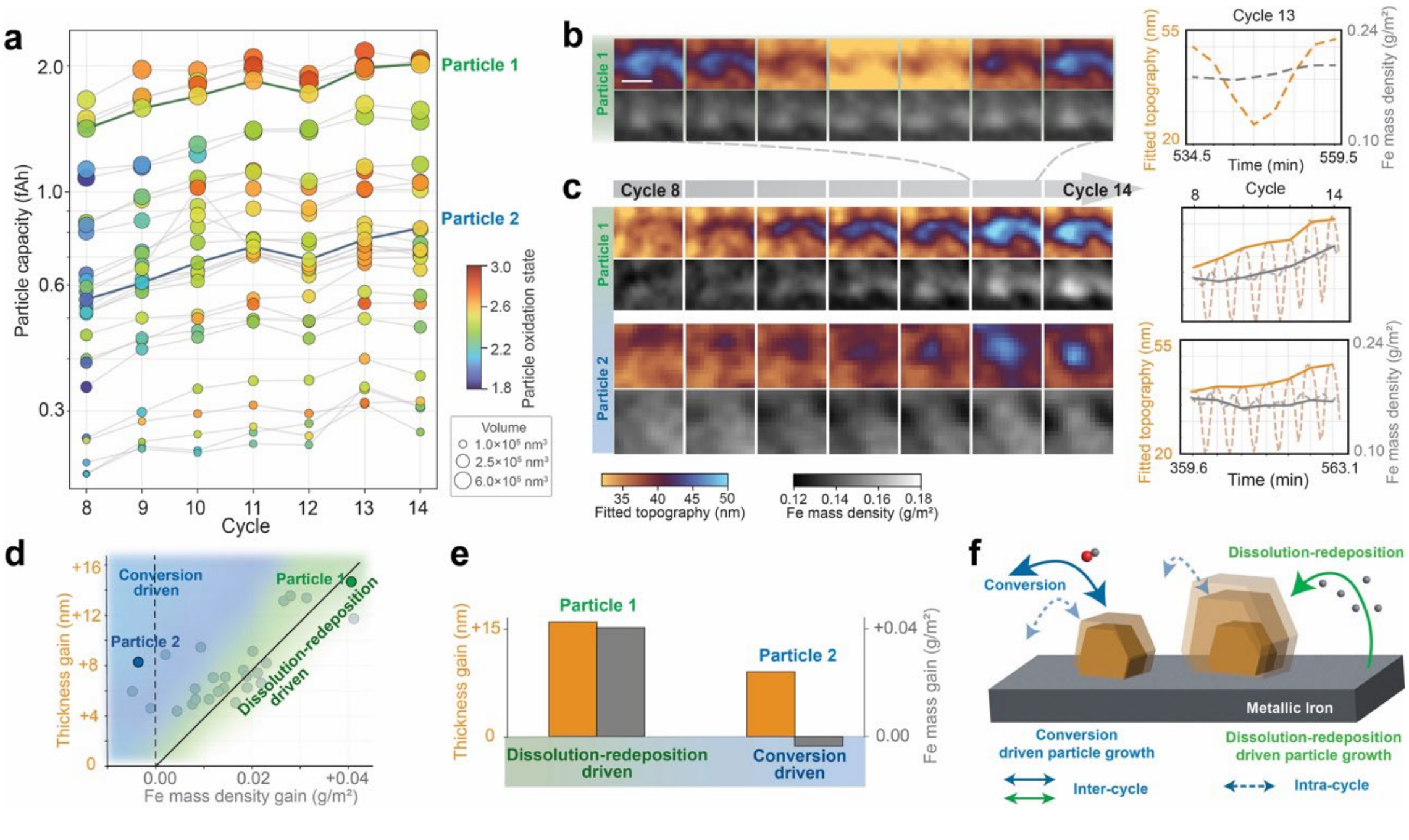

**Fig. 4 | Particle-dependent growth pathways across timescales. a**, Derived capacity trajectories of 31 particles during cycles 8–14, with particle capacity plotted on a logarithmic scale. Marker size and colour indicate absorption-derived volume and mean Fe oxidation state, respectively. **b,** Fitted-topography (top) and equivalent Fe areal mass-density (bottom) maps of particle 1 during cycle 13, with corresponding profiles at right. Reversible thickness changes

occur with little net Fe redistribution. Scale bar, 100 nm. **c,** Discharged-state maps of particles 1 and 2 during cycles 8–14, with fitted-thickness (orange) and Fe areal mass-density (grey) profiles at right. Solid lines connect discharged states and dashed traces show within-cycle variations. **d,** Thickness gain versus Fe areal mass-density gain between cycles 8 and 14 for 31 particles. The diagonal indicates the thickness gain expected from Fe accumulation under the spectral model, while excess thickness gain is consistent with conversion-related expansion. About 84% of the growing particles gain both thickness and Fe content. **e,** Net thickness and Fe areal mass-density gains for particles 1 and 2. Particle 1 gains both thickness and Fe content, whereas particle 2 thickens with little net Fe accumulation. **f,** Proposed ion-intercalation and dissolution–redeposition pathways. Solid and dashed arrows denote inter-cycle and intra-cycle processes, respectively.

## METHODS

**Fe microelectrode fabrication.** Fe microelectrodes were fabricated on prefabricated silicon nitride biasing chips (Hummingbird Scientific) by a “lift-off” process employing photolithography and electron-beam evaporation. Photolithography was performed at the Stanford Nano Shared Facilities, and Fe deposition at the Marvell Nanofabrication Laboratory, University of California, Berkeley. The chips were sequentially rinsed with acetone, isopropanol and deionized water, followed by oxygen-plasma cleaning for 5 min. SPR 3612 positive photoresist was spin-coated at 3,000 r.p.m. for 30 s and soft-baked at 90 °C for 1 min. The resist was exposed at 375 nm using an MLA150 lithography system with a dose of 90 mJ $cm^{-2}$, developed in MF-26A for 30 s and rinsed with deionized water. A 25-nm-thick Fe film was then deposited by electron-beam evaporation, followed by soaking in acetone and rinsing with isopropanol and deionized water. The Fe thickness was selected to give an optical density of about 0.75 at 707.6 eV for metallic Fe. Each microelectrode measured 10 μm × 35 μm laterally. Fe films used for electrochemical measurements and ex situ characterization were prepared using the same deposition procedure.

**Electrochemical cell and control.** Operando measurements were performed in a liquid electrochemical flow cell with a microfluidic channel between two X-ray-transparent $SiN_x$ windows. The biasing chip carried an indium tin oxide contact supporting the Fe working electrode and two patterned Pt electrodes. One Pt electrode served as the counter electrode, whereas the second remained electrically floating and acted as a local shunt. A miniature leakless Ag/AgCl reference electrode (eDAQ), positioned upstream in the electrolyte inlet, provided the reference potential. The electrolyte-path resistance between the external reference and working electrodes was adjusted to about 1 MΩ. To improve the high-frequency stability of voltage control, the Ag/AgCl reference was capacitively coupled to the floating on-chip Pt shunt through a 100 nF capacitor. This capacitive shunt reduced the long resistance–capacitance response of the electrolyte path without altering the steady-state reference potential[5]. The cell was filled with flowing, argon-saturated 0.1 M KOH, and the Fe electrode was cycled by cyclic voltammetry at 2 mV $s^{-1}$ using a BioLogic SP-300 potentiostat. Transient gas bubbles attributed to hydrogen evolution appeared during cycle 7 and between cycles 14 and 15, and were displaced shortly thereafter by continued electrolyte flow.

**Laboratory electrochemical characterization.** Separate cyclic voltammetry measurements (Extended Data Fig. 5a,b) were performed at Stanford University using a 50-nm-thick Fe working electrode in a flooded electrochemical cell (Redox.me). A Pt electrode and a leakless Ag/AgCl electrode (eDAQ) served as the counter and reference electrodes, respectively. The Fe film was cycled in argon-saturated 0.1 M KOH at 10 mV $s^{-1}$ using a BioLogic SP-300 potentiostat, beginning with a cathodic sweep. The electrolyte was purged with argon at 50 mL $min^{-1}$ for 15 min before each experiment, with purging maintained throughout cycling. Before each experiment, electrochemical impedance spectroscopy was used to determine the uncompensated resistance, which was used to correct the measured potentials for ohmic drop.

**Fast operando soft X-ray spectro-ptychography.** Experiments were performed at beamline 7.0.1.2 (COSMIC) of the Advanced Light Source, Lawrence Berkeley National Laboratory. Each three-energy stack comprised ptychographic scans acquired sequentially at 701.0, 707.6 and 709.0 eV across the Fe $L_3$ edge (Fig. 1a,c and Extended Data Fig. 1). The sample was positioned 30 μm downstream of the beam focus, producing a probe with a full width at half maximum of 1.1 μm at the sample plane (Extended Data Fig. 2). Each nominal 4 μm × 4 μm field of view was sampled on a 20 × 20 raster with a 200 nm step, maintaining sufficient overlap between adjacent illuminated regions for stable ptychographic reconstruction. Acquisition required 3.5 min per stack, yielding about 10 stacks during a typical 35-min charge–discharge cycle.

**X-ray dose estimation and reduction by sparse scanning.** The incident photon flux at 709 eV was about $\Phi_0 = 4 \times 10^8$ photons $s^{-1}$. The instantaneous absorbed-dose rate during illumination was estimated as[51],

$$\dot{D}(E) = \frac{\Phi_0 E_\gamma \left[1 - e^{-\mathrm{OD}(E)}\right]}{A_{\mathrm{eff}}\, \rho_{\mathrm{Fe}}\, t_{\mathrm{Fe}}}, \qquad (1)$$

where $E_\gamma$ is the photon energy in joules, $\mathrm{OD}(E)$ is the dimensionless optical density, $A_{\mathrm{eff}}$ is the effective illuminated area, $\rho_{\mathrm{Fe}}$ is the density of metallic Fe, and $t_{\mathrm{Fe}}$ = 25 nm is the initial film thickness. Based on the measured probe width, $A_{\mathrm{eff}}$ was approximated as 1.1 µm × 1.1 µm. This calculation normalizes the absorbed energy to the initial Fe-film mass, with 1 Gy corresponding to 1 J $kg^{-1}$. The estimated dose rate was about 5 × $10^7$ Gy $s^{-1}$ at 709 eV and below 5 × $10^6$ Gy $s^{-1}$ at 701 eV. These values describe illumination at a single probe position; cumulative dose also depends on dwell time and overlapping exposures.

Within the same approximation, the field-averaged absorbed dose per single-energy raster scan is,

$$\overline{D}(E) \approx \frac{\Phi_0 E_\gamma \left[1 - e^{-\mathrm{OD}(E)}\right] t_{\mathrm{dwell}}}{\rho_{\mathrm{Fe}}\, t_{\mathrm{Fe}} d_{\mathrm{step}}^2}, \qquad (2)$$

where $t_{\mathrm{dwell}}$ is the exposure time per scan position and $d_{\mathrm{step}}$ is the square-raster step size. Thus, for fixed photon energy, flux and sample absorption,

$$\overline{D} \propto \frac{t_{\mathrm{dwell}}}{d_{\mathrm{step}}^2} \qquad (3)$$

The highly defocused probe enabled a 200 nm raster step while maintaining sufficient overlap for ptychographic reconstruction. Relative to an 80–100 nm raster under otherwise identical conditions, this reduces the number of exposures and mean dose per scan by a factor of 4–6.25. To assess possible beam-induced material loss, we compared the repeatedly imaged region with the surrounding electrode using ptychography and wider-field spectro-STXM after cycling in the same electrolyte. No discernible preferential thinning was observed, and post-mortem SEM showed a comparable FeOOH-rich nanoplate morphology. These observations indicate that beam-induced Fe loss was limited under the reduced-dose acquisition conditions (Extended Data Fig. 6).

**Ex situ microscopy and spectroscopy.** Annular dark-field scanning transmission electron microscopy (ADF-STEM) was performed using the TEAM 0.5 microscope at the National Center for Electron Microscopy, operated at 300 kV with a probe convergence semi-angle of 25 mrad. The detector inner and outer collection semi-angles were 28 and 141 mrad, respectively. Scanning electron microscopy was performed using a Zeiss Ultra 55 field-emission microscope at the University of California, Berkeley, and a Thermo Fisher Scientific Apreo S LoVac microscope at the Stanford Nano Shared Facilities. Raman spectra were acquired using a Horiba LabRAM system with 532-nm laser excitation, and atomic force microscopy was performed in tapping mode using an Asylum Research MFP-3D instrument. Both measurements were performed at the Stanford Nano Shared Facilities

**Ptychographic reconstruction and frame alignment.** Each ptychographic scan was reconstructed independently using an automatic-differentiation-based algorithm implemented in CDTools[52]. The dataset comprised 311 three-energy stacks. Spatial resolution was estimated by fitting an edge-spread function to the amplitude profile across the reconstructed Fe-film boundary. The fitted 10–90% rise width of 34 nm was used as a conservative estimate of full-period spatial resolution (Extended Data Fig. 3), as surface roughness and gradual thickness variations at the boundary can broaden the measured edge response.

The three energy channels within each stack were first registered for spectroscopic analysis, followed by alignment across the time series to correct lateral drift and sample motion. Subpixel shifts were determined independently along the $x$ and $y$ directions using one-dimensional profiles extracted from the target and reference images. MATLAB's lsqcurvefit function was used to minimize the sum of squared differences between the shifted target and reference profiles. A displacement of $\Delta u$ pixels along $u = x\ or\ y$ was applied through a Fourier-space phase ramp

$$\tilde{T}_{\Delta u}(k) = \tilde{T}(k)\, e^{-2\pi i k \Delta u / N} \qquad (4)$$

where $\tilde{T}(k)$ is the discrete Fourier transform of the target profile, $k$ is the spatial-frequency index and $N$ is the number of sampled points. The shifted profile was obtained by inverse Fourier transformation, and the fitted shifts were applied to the corresponding reconstructed images.

**Oxidation-state mapping and effective $Fe^0$/FeOOH thickness fitting.** After spatial alignment, the transmitted intensity at each pixel and photon energy, $I(E)$, was converted to optical density using the intensity transmitted through an adjacent electrolyte-only region, $I_0(E)$, as the reference,

$$I(E) = I_0(E) e^{-\mu_l(E) d} \qquad \mathrm{OD}(E) = \ln\left[\frac{I_0(E)}{I(E)}\right], \qquad (5)$$

where $\mu_l(E)$ is the linear absorption coefficient and $d$ is the sample thickness. Optical density is dimensionless. The linear absorption coefficient is related to the mass absorption coefficient, $\mu_m(E)$, by

$$\mu_l(E) = \mu_m(E)\rho \qquad (6)$$

where $\rho$ is the mass density. At each pixel, the optical densities at 707.6 and 709.0 eV were corrected by subtracting the pre-edge value at $E_0 = 701.0$ eV. The ratio of these corrected signals was mapped to an effective Fe oxidation state by linear interpolation between reference ratios obtained from $Fe^0$, $Fe^{8/3+}$ and $Fe^{3+}$ spectra. Applying this calibration across the time series yielded pixel-resolved oxidation-state maps.

Separate metallic-Fe and FeOOH reference spectra were calibrated for effective-thickness fitting. For metallic Fe, a pattern about 25 nm thick was fabricated and its thickness measured by atomic force microscopy (AFM). Spectroscopic scanning transmission X-ray microscopy (STXM) measurements were performed on the same pattern in electrolyte under reducing bias. The spectrum was divided by the AFM-measured thickness to obtain optical density per unit thickness. At 707.6 eV, the pre-edge-subtracted value was 0.030 nm$^{-1}$, corresponding to a pre-edge-subtracted linear absorption coefficient of $3\times10^5$ cm$^{-1}$. For the FeOOH reference, a Fe film about 16 nm thick was cycled three times in 1 M KOH at 2 mV s$^{-1}$ to produce FeOOH nanoplates with a mean AFM-measured thickness of about 60 nm. The corresponding STXM spectrum was normalized by thickness in the same way, yielding a pre-edge-subtracted value of 0.012 nm$^{-1}$ at 709.0 eV. The thickness-normalized reference spectra are shown in Extended Data Fig. 1b.

For thickness fitting, the absorption at each pixel was represented as a linear combination of metallic Fe and FeOOH. Fe(II)-containing intermediates, carbonaceous radiolysis products and other phases were not included as separate components. The pre-edge-subtracted optical density was expressed as

$$\Delta\mathrm{OD}(E) = \mathrm{OD}(E) - \mathrm{OD}(E_0) \approx \Delta\mu_l^{\mathrm{Fe^0}}(E)t_{\mathrm{Fe^0}} + \Delta\mu_l^{\mathrm{FeOOH}}(E)t_{\mathrm{FeOOH}}\,, \qquad (7)$$

for $E = 707.6$ and 709 eV, where $\Delta\mu_l^j(E) = \mu_l^j(E) - \mu_l^j(E_0)$ is the pre-edge-subtracted linear absorption coefficient of component $j$, obtained from its thickness-normalized reference spectrum. Solving these two equations at each pixel yielded the effective thicknesses $t_{\mathrm{Fe^0}}$ and $t_{\mathrm{FeOOH}}$. The $Fe^{8/3+}$ reference was used only for oxidation-state assignment and was excluded from the thickness fit. Fitted topography was defined as

$$T = t_{\mathrm{Fe^0}} + t_{\mathrm{FeOOH}}\,. \qquad (8)$$

This quantity represents an absorption-derived equivalent thickness under the two-component model and should not be interpreted as a direct surface-height measurement. The equivalent Fe and O areal mass densities were calculated as

$$M_{\mathrm{Fe}}(\mathrm{g\ m^{-2}}) = 0.00787t_{\mathrm{Fe^0}}\,(\mathrm{nm}) + 0.00269t_{\mathrm{FeOOH}}\,(\mathrm{nm}), \qquad (9)$$

and

$$M_{\mathrm{O}}(\mathrm{g\ m^{-2}}) = 0.00154t_{\mathrm{FeOOH}}\,(\mathrm{nm}). \qquad (10)$$

These are model-dependent estimates of elemental content per unit projected area. The oxygen estimate includes only oxygen assigned to the fitted FeOOH component.

**Treatment of Fe(II) in the spectral model.** The effective-thickness analysis represents the absorption at each pixel as a linear combination of metallic Fe and FeOOH. Operando Fe $L_3$-edge spectra and ex situ Raman measurements support these as the dominant endmembers during cycling. Raman spectra show δ-FeOOH and α-FeOOH during the first five cycles, with α-FeOOH predominating thereafter (Extended Data Fig. 5c). The band near 660 $cm^{-1}$ may also contain a contribution from magnetite, $Fe_3O_4$, so minor Fe(II)-containing or mixed-valence species cannot be excluded. Although no distinct bulk Fe(II)-rich phase was resolved by the available ex situ measurements, Fe(II)-containing intermediates may participate transiently in alkaline Fe redox chemistry[43–47].

STXM and ptychography integrate absorption through the film thickness, so minor surface contributions may be difficult to distinguish from the dominant bulk signal. Sequential acquisition of each three-energy stack takes 3.5 min, limiting the ability to resolve short-lived intermediates. At a sweep rate of 2 mV $s^{-1}$, this corresponds to a potential change of about 0.42 V during a monotonic sweep. The thickness fit also contains only two pre-edge-corrected spectral constraints at 707.6 and 709.0 eV. A three-component $Fe^0$/Fe(II)/Fe(III) thickness fit would therefore be underdetermined without additional photon energies or external constraints. We used a two-component metallic-Fe/FeOOH model, in which unresolved Fe(II) contributions can affect the fitted component thicknesses. The resulting thicknesses, fitted topography and equivalent areal mass densities are consequently model-dependent estimates.

For oxidation-state mapping, we used $Fe^0$, $Fe^{8/3+}$ and $Fe^{3+}$ reference spectra. The $Fe^{8/3+}$ reference corresponds to the mean Fe valence in $Fe_3O_4$ and provided an intermediate-valence calibration because a reliable Fe(II) standard could not be obtained. This reference was excluded from the thickness fit, and its use does not establish the presence of magnetite in the electrode. Extended Data Fig. 6d compares oxidation-state maps calculated with and without this reference. The spatial mean ± standard deviation was 2.33 ± 0.38 with the reference and 2.30 ± 0.35 without it, a difference of less than 5% in the mean. This comparison indicates that the mean mapped oxidation state is only weakly sensitive to the calibration choice; it does not quantify the effect of unresolved Fe(II) on the thickness or areal mass-density estimates.

**Imaging-derived capacity calculation.** The areal capacity of each charge or discharge half-cycle was estimated from the change in effective FeOOH thickness obtained from the two-component spectral fit. We defined $\Delta t_{\mathrm{FeOOH}}$ as the magnitude of the thickness difference between the beginning and end of the half-cycle. The calculation assumes electrochemical conversion between metallic Fe and FeOOH, neglecting contributions from chemical oxidation, X-ray-induced reactions and dissolved Fe species. The resulting values are model-dependent capacity estimates derived from imaging, rather than independent electrochemical measurements.

The net conversion between $Fe^0$ and $Fe^{3+}$ in FeOOH transfers three electrons per Fe atom. The areal capacity was calculated as

$$Q_{\mathrm{A}} = \frac{nF\rho_{\mathrm{FeOOH}}\Delta t_{\mathrm{FeOOH}}}{3.6M_{\mathrm{FeOOH}}}, \qquad (11)$$

where $Q_A$ is the areal capacity in mAh m$^{-2}$, $\rho_{FeOOH}$ is the FeOOH density in kg m$^{-3}$, $M_{FeOOH}$ is its molar mass in kg mol$^{-1}$, $\Delta t_{FeOOH}$ is expressed in metres, $n = 3$ and $F = 96{,}485$ C mol$^{-1}$. Division by 3.6 converts coulombs to milliampere-hours. Using the FeOOH density adopted in the spectral model gives

$$Q_A\ (\text{mAh m}^{-2}) = 3.869\ \Delta t_{FeOOH}\ (\text{nm}) \qquad (12)$$

The calculation was applied pixel by pixel to generate spatially resolved capacity maps. FeOOH thickness increases during discharge and decreases during charge were both reported as positive capacity magnitudes.

**Particle segmentation and tracking.** Particle-like features were identified from the fitted-topography maps using a semi-automated workflow. Within a predefined region of interest, local maxima were detected independently along image rows and columns, and their lateral extents were combined to define candidate particle regions. Morphological operations removed isolated noise and weak connections between neighbouring features. Touching particles were separated using a watershed transform applied to the distance map. Regions were excluded if they contained fewer than 10 pixels (1,440 nm$^2$), touched the image boundary or failed the intensity criterion. Each retained connected region was assigned a unique particle label. After spatial and temporal registration, the particle mask defined from a selected reference frame was applied unchanged to all analysed charged and discharged frames. This fixed-mask approach maintained consistent particle-centred regions throughout cycling and avoided variability introduced by repeated segmentation. The resulting measurements therefore quantify changes within predefined regions without tracking evolving particle boundaries.

**Acknowledgements** This work was primarily supported by the U.S. Air Force Office of Scientific Research through the Multidisciplinary University Research Initiative under Award No. FA9550-23-1-0281 for the development of low-dose imaging methods, and by the Gordon and Betty Moore Foundation under Grant No. 10784 for hyperspectral imaging. J.M., Y.C. and D.J. acknowledge partial support from the Air Force Office of Scientific Research under Award No. FA9550-25-1-0334. This work was also supported in part by STROBE, a National Science Foundation Science and Technology Center, under Award No. DMR-1548924. This research used resources of the Advanced Light Source, a U.S. Department of Energy Office of Science User Facility under Contract No. DE-AC02-05CH11231. Ptychography experiments were performed at beamline 7.0.1.2 (COSMIC) of the Advanced Light Source. Part of this work was conducted at the Stanford Nano Shared Facilities, the Stanford Nanofabrication Facility and the Marvell Nanofabrication Laboratory at the University of California, Berkeley. E.Z.C. was supported in part by an Advanced Light Source Doctoral Fellowship in Residence.

**Author contributions** J.M., W.C. and D.S. supervised the project; X.Z. and E.Z.C. redesigned the electrochemical cell and set up the operando experiments; X.Z., E.Z.C. and A.B. fabricated the Fe nanopatterns;

X.Z., Y.C., E.Z.C., A.B., D.J., H.Z., L.T. and F.C. carried out the soft X-ray spectro-ptychography experiments; A.D. and D.S. prepared the beamline and assisted with experiments. Y.C., X.Z. and J.M. performed image reconstruction and analysis. X.Z., Y.C., E.Z.C., A.B. and J.M. analysed the data and interpreted the results; H.S. and X.Z. performed ADF-STEM, A.B., X.Z., Y.S. and E.Z.C. conducted electrochemical tests and other ex-situ characterization; X.Z., Y.C., J.M. and E.Z.C. wrote the paper. All authors commented on the paper.

**Competing interests** The authors declare no competing interests.

**Additional information**

**Correspondence and requests for materials** should be addressed to J.M. (j.miao@ucla.edu).

**Data availability**

All data supporting the findings of this study will be deposited in a public repository and made available upon publication of this article.

**Code availability**

Ptychographic reconstructions were performed using the publicly available CDTools software (https://github.com/cdtools-developers/cdtools). Custom code developed for image registration, chemical-state mapping, quantitative analysis and particle tracking will be made publicly available on GitHub upon publication of this article.

1. Rieb, C., Leclerc, N., Méry, S., Hébraud, A. & Swaraj, S. Estimating spatial resolution and X-ray radiation dose in a comparative study of composite organic nanoparticles using soft X-ray scanning transmission X-ray microscopy and soft X-ray ptychography. *J. Phys. Chem. C* **129**, 18537–18547 (2025).
2. Levitan, A., Cain, M. & Kutakh, A. *CDTools* (Version 0.2.0). GitHub (2024): github.com/cdtools-developers/cdtools.